 \documentclass[referee]{aa} % for a referee version
 
\begin{document}
\title{Memberships  and  CM  Diagrams  of  the  Open
                     Cluster  NGC$\,$7243}%\thanks{Table 8 is only available}
%           in electronic form at the CDS via {\tt http://cdsweb.u-strasbg.fr/.....}}
\author{E.G. Jilinski\inst{1,2,3} \and V.N. Frolov \inst{1}
 \and J.K. Ananjevskaja \inst{1} \and J. Straume\inst{4}
 \and N.A.~Drake \inst{2,5}
}
\offprints{E.G.Jilinski, Observat\'orio Nacional/MCT, Rua Gal. Jose
   Cristino 77, S\~ao Cristov\~ao, Rio de Janeiro, Brazil.
   e-mail: jilinski@on.br}

\institute{Main Astronomical Observatory, Pulkovo, St. Petersburg, Russia
\and
       Observat\'orio Nacional/MCT, Rio de Janeiro, Brazil
\and
      Laborat\'orio Nacional de Computa\c c\~ao Cient\' \i fica/MCT, Petropolis,
       Brazil
\and
       Radioastrophysical Observatory, Baldone, Riga, Latvia
\and
       Sobolev Astronomical Institute of St.~Petersburg State University,
       St.~Petersburg, Russia}
\date{ }

\authorrunning{Jilinski et al.}

\titlerunning{Open Cluster NGC$\,$7243}

 %______________________________________________________________________________
 %
\abstract{
 The results of astrometric and photometric investigations of the open
 cluster NGC$\,$7243
 are presented. Proper motions of 2165 stars with  root-mean-square
 error of $1.1\;{\rm mas\,yr^{-1}}$ were obtained by means of PDS scanning of
 astrometric plates
 covering the time interval of 97 years.         A total of 211 cluster members
 down
 to $V=15.5\;$mag have been identified. $V$ and $B$ magnitudes have been
 determined for 2118 and 2110 stars respectively.
 Estimations of mass
 ($348 {M_\odot} \le M_{TOT} \le 522 {M_\odot}$),
 age $(t=2.5\cdot 10^8\,$yr),
 distance $(r=698\,$pc) and  reddening $(E_{B-V}=0.24)$ of the cluster
 NGC$\,$7243 have been made.
 \keywords{open clusters and associations: general -
                open clusters and associations: individual:  NGC$\,$7243
            }
 }

 %%\end{abstract}

 \maketitle

 %______________________________________________________________________________

 %______________________________________________________________________________

\section{Introduction}

 The open cluster NGC\,7243 is located in the Lacerta constellation.
 It has the following coordinates:
 $\ell=98\fdg 9$, $b=-5\fdg 6$, $\alpha=22^{\rm h} 15^{\rm m}$,
 $\delta=+49\degr 53\arcmin$ (2000.0).
 In the Lyng\aa\, (1987)                  %(\cite{lynga})
 catalogue it is classified as Tr\"umpler (1930) class $\,$II$\,$2m
 with %a total amount of
 40 cluster members and an angular diameter of $21\arcmin$.
 Early investigations of this cluster led Raab (1922)         %(\cite{raab})
 to the conclusion that, in fact, it consisted of two clusters,
 but a lack
 of sufficient observational data made it impossible to check this statement at
 that epoch.
 Zug (1933)           %(\cite{zug})
 was the first investigator to study the interstellar absorption in the cluster
 area.
 He denoted the brightest stars and these numbers are still used in some
 papers. Relative proper motions of 814 stars in the cluster area were
 for the first time obtained by Lengauer (1937)            %(\cite{lengauer})
 from  plates taken with the Normal
 Astrograph of the Pulkovo Observatory with a 33 years mean epoch difference.
 As a result, 32 cluster members were selected. Later
 Rahmatov and Muminov (1985)                        % (\cite{rahmatov})
 determined proper motions of 1287 stars in the region of NGC\,7243 using the
 plates
 of the Normal Astrograph in Tashkent (Uzbekistan)
 as the first epoch plates.
 Due to intensive urban light
 pollution, the second
 epoch plates had to be taken with the 40\,cm Zeiss Astrograph of the
 Kitab Latitude Station (Uzbekistan).
 In the circle of $40\arcmin$ radius,
 28 possible cluster members with a membership probability of more
 than $50\%$ were selected.
 van Schewick (1957) using only one pair of plates          %(\cite{schewick})
 confirmed the existence of  cluster members, previously found by Lengauer,
 but did not add any one new.
 Unfortunately, none of the above mentioned  investigations, for one reason or
 another (absence of photometric data, small epoch difference, small number of
 plate pairs, etc.)
 permitted  to select all probable  members of the cluster NGC\,7243 in the
 observed magnitude range and to estimate membership probabilities.

 The Main Astronomical Observatory at Pulkovo has 19 first epoch
 plates (obtained from year 1897 to 1949). This allows to solve the astrometric
 part of the problem with a high degree of confidence.
 For a reliable segregation of the cluster members it is necessary to use 
 photometric data based on the international photometric system.

 Photometric plates of the NGC\,7243 region
 in $B$ and $V$ passbands  were obtained with the Schmidt Telescope (80/120/240)
 of the Latvian Radioastrophysical Observatory.

 \section{Observations}

 The  observational  material  used  for  the  determination  of  the
 relative  proper motions  in  the field of  NGC$\,$7243  is  shown  in
 Table~1.  All plates  are  from  the  Normal  Astrograph
 Collection  of  Pulkovo  Observatory.
 The Pulkovo  Normal  Astrograph  has a diameter of 33 cm, and a scale  of
 ${\rm 60\arcsec\,mm^{-1}}$.
 All plates are in the $m_{pg}$ passband and
 have  been obtained with exposition times listed in Table~1.
 The  plate  and  film  material  used  for  the $BV$  photometry  is
 shown  in  Table~2.

 \begin{table*}
       \caption[]{Astrometric plate  material}
 \begin{tabular}{lcll}
 \hline
  Plate         &   Exposure      & \,\,\,Epoch     &        Quality\\
           &    (min)     &         &        \\
 \hline
 \multicolumn{4}{c}{Early  epoch}\\

 A  202  &              20 &              1897  Aug  13   &      poor\\
 A  346  &              30 &              1899  Sep   5   &      good\\
 A  525  &              62 &              1901  Sep  14   &      good\\
 A  708  &              60 &              1903  Sep  24   &      poor\\
 A  976  &              22 &              1907  Sep  11   &      high\\
 B  261  &              30 &              1910  Oct  11   &      good\\
 C  995  &              30 &              1933  Nov  23   &      high\\
 C  996  &              40 &              1933  Nov  23   &      high\\
 C  997  &              30 &              1933  Nov  23   &      high\\
 K   11  &              80 &              1934  Feb   9   &      high\\
 D  232  &              21 &              1949  Oct  30   &      good\\
 \multicolumn{4}{c}{Recent  epoch}\\
   15453  &              10 &              1987  Oct  18   &      good\\
   15454  &              10 &              1987  Oct  18   &      good\\
   17670  &              10 &              1995  Oct  15   &      high\\
   17671  &              20 &              1995  Oct  15   &      high\\
   17696  &              10 &              1996  Jan  24   &      good\\
   17697  &              10 &              1996  Jan  24   &      high\\
   17704  &              10 &              1996  Jan  25   &      poor\\
 \hline
 \end{tabular}
 \end{table*}

 \begin{table*}
       \caption[]{Photometrical plate material}
 \begin{tabular}{ccccc}
 \hline
    Plate      &            Epoch       &     Passband   &  Photomaterial    &  Quality\\
 \hline
       18575   &       1991  Jan  13/14       &  $V$  &     A600 H +  YG 17  &     high\\
       19026   &       1992  Jan  25/26       &  $V$  &     A600 H +  YG 17  &     high\\
       19027   &       1992  Jan  25/26       &  $V$  &     A600 H +  YG 17  &     high\\
       19202   &       1992  Aug  27/28       &  $V$  &     A600 H +  YG 17  &     high\\
       18536   &       1990  Dec  16/17       &  $B$  &      ZU 21 +  GG 13  &     high\\
       19028   &       1992  Jan  25/26       &  $B$  &      ZU 21 +  GG 13  &  $\,$good\\
       19230   &       1992  Dec  17/18       &  $B$  &      ZU 21 +  GG 13  &     high\\
 \hline
 \end{tabular}
 \end{table*}

 \section{Astrometry}

 The deepest plate (K$\,$11) was measured by means of the semi-automatic
 coordinate measuring machine ``Askorecord'' with a precision of about 0.01 mm.
 Serial numbers from 1 up to 2623 were assigned to all stars in a square
 size of $80\arcmin\times 80\arcmin$ centered on the cluster.
  These positions were used as preliminary ones when all plates were scanned
 with the Observat\'orio Nacional (Brazil) PDS$\,$1010 microdensitometer.
 The stellar images were analyzed  with a $10\times 10\;\mu$m diaphragm.
 The plates were scanned in two orientations with a 180 degrees rotation
 to eliminate  possible errors of the image treatment and to minimize the
 influence
 of systematic errors of the PDS microdensitometer. The analysis and
 comparison of the direct and reverse measurements permitted to evaluate the
 random and systematic errors of
 measurements as $\pm 1.5\;\mu$m in both coordinates.

 The star image positions were determined through two-dimensional elliptical
 Gaussian fits to the matrix of pixels of the image. These  calculated positions
 of the stars do not depend on the orientation of the scanning.

 The  method  used for proper motion determination traditionally
 suggests  compiling of plate pairs.
 The combinations of plates that form such pairs, are listed  in Table~3.

 \begin{table*}   %Tabl.3
       \caption[]{Plate pairs used for the proper motions determination}
 %\begin{center}
 \begin{tabular}{cccccc}
 \hline
   Pair  &  Plates   &   Epoch  &   Pair &  Plates  & Epoch\\
         &           &   difference  &      &   &     difference\\
         &           &     yr        &       &               &  yr  \\
 \hline
   1  &   A$\,$346-17696 &   97 &    6  &   C$\,$995-17671   &      62\\
   2  &   A$\,$525-17697 &   95 &    7  &   C$\,$996-17670   &      62\\
   3  &   A$\,$708-15453 &   84 &    8  &   C$\,$997-17671   &      62\\
   4  &   A$\,$976-15670 &   88 &    9  &   $\;$K$\,$11-17696  &    62\\
   5  &   B$\,$256-15454 &   77 &   10  &   D$\,$232-17697   &      47\\
 \hline
 \end{tabular}
 %\end{center}
 \end{table*}

   The  mean  epoch  difference is 74  years. Some of the second epoch plates
 were used in the pairs twice. This was caused  by unfavorable conditions
 of observation at Pulkovo Observatory due to  urban lights.
 The plate $17704$ was excluded because of poor quality of
 measurements. Besides, nine plates of the first epoch being not deep enough
 were not used.

    The positions of stars on all plates were related to one conventional
 center (star 898 in our catalogue). The plates of the second epoch
 were used as plates of comparison, therefore the coordinate axes on the plates
 were rotated so that the axis $Y$ was oriented along the sky meridian.
 For this purpose stars from the catalogue AGK$\,$3 were used.

 The reference stars were chosen in the $B$ magnitude interval between $13.5$
 mag and $14.5$ mag.
 To avoid  the so-called ``cosmic''  error, i.e. the position change of a
 system of reference stars because of their real movements, the following
 operation was applied. All the plates used were transferred on one - D\,232,
 as far as its epoch is median for all others. Then for each
 reference star candidate, graphics were considered, where on abscissa  the
 epoch of plates and on ordinate - coordinate $X$ or $Y$ of the star were
 plotted.
 Further, a straight line was drawn  by  least square fitting.
 The star was included in the reference stars list, only if this straight
 line was parallel to the coordinate axis, which testified the absence of an
 appreciable
 proper motion.

    Thus, 99 reference stars (of average $B$ magnitude equal to 14.04) existing
 on
 all plates and uniformly distributed in the studied region, except the cluster
 area itself, were selected.

    Quadratic polynomial approximations were used to match the position of the
 star on the the 1-st epoch plate  to that of the 2-nd epoch:
 $$        ax + by + cxy + dx^2 + ey^2 + f =\Delta x\eqno (1) $$
 $$       a'x + b'y + c'xy + d'x^2 + e'y^2 + f' = \Delta y  $$
 The mean errors of stars relative proper motions for each plate pair
 are given in Table~4.

\begin{table*}[h]  %Table  4
 \caption[]{Errors of stars relative proper motions obtained using different
     plate pairs}
 \begin{tabular}{cccc}
 \hline
 Plate pair & Number  & $\epsilon _x$ & $\epsilon _y$ \\
        & of stars &  \multicolumn{2}{c}{(mas\,y$r^{-1}$)}   \\
 \hline
     1  &    2025  &    3.37  &    2.73\\
     2  &    1923  &    3.05  &    2.62\\
     3  &    2121  &    2.77  &    2.34\\
     4  &    2140  &    2.42  &    2.08\\
     5  &    2122  &    2.94  &    2.65\\
     6  &    2026  &    3.37  &    3.25\\
     7  &    2084  &    3.15  &    2.86\\
     8  &    2107  &    3.30  &    2.92\\
     9  &    1956  &    3.99  &    3.29\\
 $\!\!\!10$ &1716  &    4.33  &    4.13\\
 \hline
 \end{tabular}
\end{table*}

 In calculating average values of $\mu _x$ and $\mu _y$, weights of
 individual plate
 pairs were not  taken into account since the difference between the mean
 and weighted mean  do not exceed the accidental errors.
 Values of the estimated root-mean-square\ (rms)  errors of the obtained proper
 motion  errors do not depend
 on the position of a star in the studied field, but increase 1.5 times from
 bright to faint stars. The mean rms errors of the proper motion components for
 all observed magnitudes are the  following:
 $$ \sigma _x = \pm 1.09\;{\rm mas\,yr^{-1}},\;\, \sigma _y =\pm 0.96\;{\rm
 mas\,yr^{-1}}.  $$
 In the whole, relative proper motions were determined for 2165 stars in the
 investigated region.

 \section{Photometry}

 Photographic $BV$ photometry of the stars in the cluster area was made using
 the material described in Section 2.
 The measurements of the plates were performed
 by means of the microphotometer ``Sartorius'' with an iris diaphragm.
 The fluctuations of the zero-point of the scale of the device during the
 measurements
 were checked and introduced into instrumental magnitudes  as  correspondent
 corrections. The measured area is characterized        by the existence of a
 large amount of  weak stars forming a rather irregular background.
 For this reason, more than 10\% of all stars of the catalogue are  either
 blended or fall simultaneously in the field of view of the device.
 Besides this, in the diaphragm, together  with a measured star, often enter
 one or several weak stars that are not numbered.
 Nevertheless, the measurements of such stars were done. Stellar
 magnitudes  obtained this way are approximate as their errors
 are as high as $0.5$ mag. In the final catalogue
 they are denoted by an attribute ``:''. Other doubtful cases, for example,
 the absence of a star on one of the plates and etc. are marked by colon too.

 Usually in measurements with iris-photometers, the background of a plate is
 considered as regular and is not specially taken into account. But for the
 field of NGC$\,$7243 it is not so, mainly in the neighbourhood of bright stars,
 of which, at the first sight, the cluster consists.
  That is why each plate was measured twice: with a consideration of the
 background and without it. The comparison of the results
 showed that  consideration of the
 background improves a convergence of magnitude values obtained
 with different plates not more than for 0.02 mag, considerably less
 than the photometric  error ($0.05$ mag). However, in the final catalogue
 photometry
 obtained with the accounting of the background is given using
  the well-known   formula of Weaver (1962)     %(\cite{weaver}):
       $$ d^2=D^2-\left(d_f^2-d_{sf}^2\right),\eqno (5)  $$
 \noindent where  $D$ -  aperture reading at prompting on a star,

 \noindent $d_f$ - aperture reading for an average background near the star,

 \noindent $d_{sf}$ - aperture reading for an average standard background,

 \noindent $d$  - aperture reading for a star, related to an average standard
                background.
 The background near each star was measured from two up to four times with
 subsequent averaging.

 The photometric calibration of every photometric plate was done using
 photoelectrically
 measured $B$ and $V$ magnitudes of stars in the region of  NGC\,7243,
 published in two works:
 Mianes \& Daguillon (1956)  and   %(\cite{mianes})
 Hill \& Barnes (1971)    %(\cite{hill})
 (hereafter referred to as MD and HB respectively).
In the first work the number of stars is 39, in the second - 20, with 9 
common stars. Since there is a systematic difference between these two
 sets of the standards, the work of MD was used as the principal and
  the values $B$ and $V$ from the work of HB were reduced to its system.
 Three stars from the GSC (Lasker et al. 1988)
 %\cite{lasker})
 were added to this list.
 The full list of the photoelectric standards is  given in Table~5.

 \begin{table*}  %Table 5
       \caption{Photoelectric photometric standard stars in the region of
 NGC\,7243}
 \begin{tabular}{rcccrcrcccrc}
 \hline
 This & Len-  & $V$ &${B-V}$&${U-B}$& Notes & This & Len-  &
 $V$&${B-V}$&${U-B}$&
 Notes\\
 paper & gauer &     &     &     &       & paper & gauer &    &     &     & \\
 %N$^\circ$  & N$^\circ$ &     &     &     &       & N$^\circ$ & N$^\circ$ &  & &     &      \\
 \hline
   912 & 475 &  8.43 &  0.11 & $-0.20$ &      &  869 &  382 & 11.35 &  0.64  &  0.16  &   *  \\
   909 & 480 &  8.74 &  0.12 & $-0.22$ &    P &  892 &  489 & 11.44 &  1.44  &  1.38  &      \\
   757 & 482 &  9.07 &  0.14 & $-0.23$ &      &  743 &  476 & 11.55 &  0.33  &  0.20  &      \\
   879 & 377 &  9.21 &  0.08 & $-0.30$ &    * &  761 &  485 & 11.68 &  0.29  &  0.22  &      \\
  1085 & 380 &  9.22 &  0.09 & $-0.22$ &      &  775 &  497 & 11.22 &  0.27  &  0.19  &      \\
   895 & 362 &  9.25 &  0.15 & $-0.23$ &    * & 1086 &  379 & 11.74 &  0.25  &  0.19  &      \\
   902 & 358 &  9.35 &  0.13 & $-0.21$ &    Q & 1097 &  384 & 11.79 &  0.26  &  0.18  &      \\
   896 & 361 &  9.60 &  0.12 & $-0.26$ &    * &  897 &  363 & 11.79 &  0.26  &  0.16  &      \\
  1078 & 370 &  9.93 &  0.05 & $-0.26$ &      &  881 &  372 & 11.80 &  0.23  &  0.18  &      \\
   771 & 490 & 10.10 &  0.12 & $-0.26$ &      &  762 &  488 & 11.82 &  0.62  &  0.18  &    T \\
   898 & 360 & 10.27 &  1.75 & $\,$2.10 &     & 1068 &  364 & 11.93 &  0.39  &  0.23  &      \\
  1099 & 386 & 10.27 &  0.15 & $-0.06$ &  R * & 1084 &  378 & 12.08 &  0.27  &  0.20  &    * \\
   776 & 496 & 10.39 &  0.13 & $-0.08$ &      &  781 &  502 & 12.19 &  0.42  &  0.30  &      \\
   908 & 350 & 10.40 &  0.45 &  $\,$0.20 &    &  907 &  355 & 12.20 &  0.72  &  0.29  &      \\
   880 & 373 & 10.50 &  0.62 &  $\,$0.16 &  * &  778 &  501 & 12.21 &  0.34  &  0.24  &      \\
   876 & 510 & 10.54 &  0.19 &  0.00    &     &  922 &  338 & 12.29 &  0.40  &  0.28 &      \\
  1081 & 374 & 10.66 &  0.72 &  $\,$0.27 &    &  888 &  366 & 12.30 &  0.45  & 0.36 &      \\
   768 & 493 & 10.73 &  0.21 & $-0.05$ &      & 1090 &  376 & 12.37 &  0.44  & 0.28 &    * \\
   924 & 471 & 10.76 &  0.20 & $-0.01$ &      &  905 &  483 & 12.41 &  0.48  & 0.29 &      \\
  1104 & 392 & 10.86 &  0.13 &   0.00   &      &  875 &  511 & 12.42 &  0.37  & 0.26 &      \\
  1075 & 368 & 10.89 &  0.14 & $-0.01$ &      &  877 &  375 & 12.50 &  0.46  & 0.26 &      \\
   874 & 513 & 11.08 &  0.70 &  $\,$0.25 &  S * &  772 &  494 & 12.66 &  0.41  & 0.28 &      \\
   759 & 484 & 11.15 &  0.22 &  $\,$0.05 &      &  792 &  515 & 12.69 &  0.42  & &   G  \\
   904 & 487 & 11.20 &  0.18 &  $\,$0.01 &   *  &  793 &      & 12.77 &  1.29  & &   H  \\
   780 & 503 & 11.24 &  0.29 &  $\,$0.21 &   *  &  884 &  505 & 12.83 &  0.25  & $-0.02$ &      \\
   886 & 369 & 11.34 &  0.19 &  $\,$0.11 &   *  &  918 &      & 15.04 &  0.73  & &   I  \\
 \hline
 \multicolumn{12}{l}{\footnotesize 1) * - the star merging in blend or close
 contact
          with  weak star(s) without  number}  \\
 \multicolumn{12}{l}{\footnotesize 2)  By  capital letters are denoted the stars
        of the  photometric sequence of the catalogue GSPS (plate P-188)} \\
 %\multicolumn{12}{l}{\footnotesize $\;\;\;\;$ of the catalogue GSPS,} \\
 \multicolumn{12}{l}{\small 3) For the star 886 in the catalogue of MD faulty
 photometry  is given} \\
 \multicolumn{12}{l}{\small 4) The star 872 (512) is not included in the list,
 since its $V$ magnitude determined by us is one} \\
 \multicolumn{12}{l}{\small  $\;\;\;\;$  magnitude brighter than that given in
 MD             (MD: $V= 12.62$ mag, $B-V = 0.35$ mag; the authors} \\
 \multicolumn{12}{l}{\small $\;\;\;\;$ catalogue: $V = 11.52$ mag, $B-V = 0.24$
 mag)} \\
 \end{tabular}
 \end{table*}

About 30 stars spreading regularly in the whole range of  magnitudes
were used for plotting  characteristic curves.
Unfortunately, the number of photoelectric standards  for faint stars is
extremely low, and that results in an almost double  increasing of the error.
The characteristic curves for every photometric plate were fitted by
polynomials.
As an example of the calibration, a characteristic curve for the plate 19026
is presented in Fig.~1.

\begin{figure} [t]
 \vspace {8cm}
 \caption []{Characteristic curve for the plate 19026.
 I(x) are readings of the iris microphotometer ``Sartorius'' and
 $B$ are  corresponding photoelectric stellar magnitudes from Table 5.}
 \vbox{\includegraphics{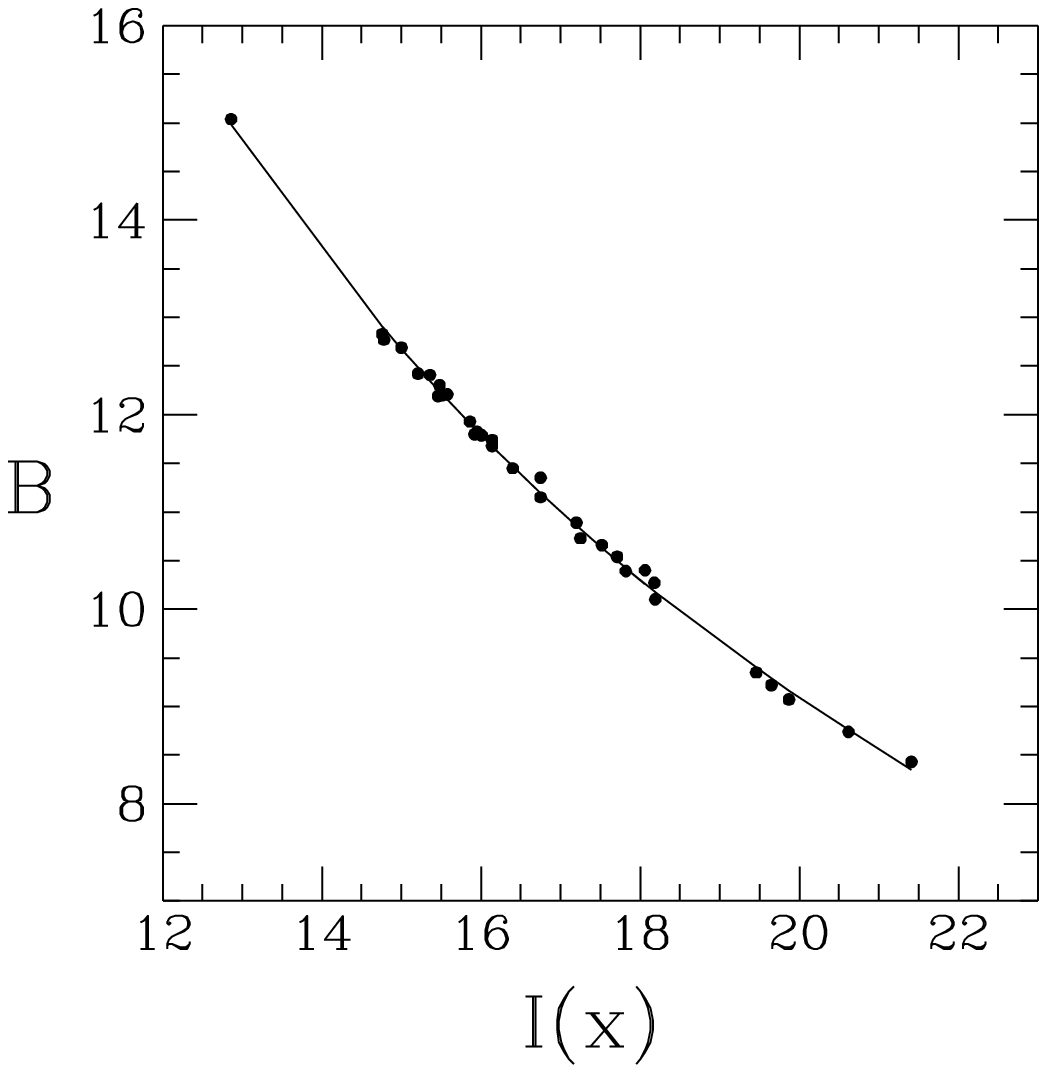}}
\end{figure}

Instrumental magnitudes were transformed into  $B$ and $V$ taking
into account the colour equation (CE):
$V$ magnitude  varies on $0.05$ mag,  while $B-V$ changes from $-0.20$ to 2.11,
for the $B$ magnitude this variation reaches $0.15$~mag.
These CEs were used for every $B$ or $V$ plate transformation.
The resulting $B$ and $V$ magnitudes are free of CE.

The external rms errors of our photometry calculated
relative to the corresponding photoelectric values for 39 stars are the
following:
 $$  \sigma _V = \pm 0.05\;{\rm mag},\;\, \sigma_{B-V} = 
     \pm  0.07\;{\rm mag}. $$
 The internal rms errors were calculated only for stars with  reliable
 photometry. Blended stars or stars contacted with not denoted ones, as well
 as stars that had not been measured, for some reason or another, on all
 plates obtained  in the same passband  were excluded.
 Finally,  photometric magnitudes $B$ and $V$ were obtained  for
 2110 and 2118 stars  respectively.
 Photometry errors increase slightly  from the center of the plate to its edges.
 The following values of the internal rms errors of the photometry were adopted:
 $\sigma_V = \pm 0.04\;{\rm mag}, \;\, \sigma_B = \pm 0.06\;{\rm mag}, \;\,
 \sigma_{B-V} = \pm 0.07\;{\rm mag}.$

 \section{Members segregation}

 Proper motions used for open cluster membership segregation should be free of
 any systematic errors. The most important systematic error is the magnitude
 equation (ME) or the dependence of the observed star positions and the derived proper
 motions on stellar magnitudes.

If all observations were done with the same telescope and with a similar type
of astroplates, under similar observational conditions (hour angles, zenith 
distances, coordinates of astroplate centers etc.) all star positions observed 
at different epochs would be contaminated with nearly the same  systematic errors. 
The resulting proper motions, calculated as the difference of star positions at 
different epochs, would be free of the principal systematic errors.

The methods of ME estimation used in Pulkovo Observatory were described in details 
by Lavdovsky (1961). All proposed methods assume preliminary
separation of cluster members from background stars and are based either on the 
analysis of proper motions of background stars only (first method), or of
probable cluster members only (second method).

In the first method, if we  exclude from consideration the most dense part of the VPD,
we exclude the main part of probable cluster members. At the same time, stars
with very high proper motions ($> 50 \;{\rm mas\,yr^{-1}}$) should be eliminated too.
The residual stars are (mostly) the background stars and their proper motions reflect
the solar motion to the Apex. The background stars are distributed randomly in
a space, so fainter stars are, on average, situated at larger distances from
the Sun. 
Thus, the parallactic displacements caused by solar motion are smaller for
fainter stars. These statistical parallactic displacements were calculated for
different stellar magnitude range using the tables published by Zhukov (1966). 
When these statistical  parallactic displacements are eliminated from the 
observed proper motions, the  residual dependence on stellar magnitude presents the ME. 
The obvious disadvantage of this method is a very low precision of the 
secular parallaxes determined till now. 
At best, this method permits to estimate the linearity and the slope 
of ME for all observed stellar  magnitudes. 

In the second method, the estimation of the ME was based on
the proper motions of cluster members preliminary selected by radial velocities. 
In Table 6 we collected all stars with known radial velocities in
the cluster region. Data from HB and Barbier-Brossat \& Figon (2000)
were used. 
Firstly, we excluded from the consideration the stars known to be 
spectral binaries and determined the mean radial velocity ($\overline{V}_{\rm rad}$)
and its standard deviation ($\sigma$). 
After that,  all stars with the radial velocities
$\mid\! V_{\rm rad} - \overline{V}_{\rm rad}\!\mid \, > \sigma$ were excluded too. 
Thus, we obtained the mean cluster radial velocity of $-11.1 \pm 1.2$ km\,s$^{-1}$ 
for 22 stars marked with symbol $\dagger$ in Table 6. 
This value was accepted as the mean radial velocity of the cluster.
If we consider that these 22 stars are real cluster members, they should have 
the same values of proper motions. But, in reality, the dependence of the observed 
proper motions on stellar magnitude was found for these 22 stars which
permitted us to estimate the ME as 
$+0.59 \pm 0.07\;{\rm mas\, yr^{-1}\, mag^{-1}}$ for
 $\mu_x$, and $0.00 \pm 0.05\;{\rm mas\, yr^{-1}\, mag^{-1}}$ for $\mu_y$.
However, because radial velocities were determined for a small number 
of bright stars ($B \le 12.5$ mag)
only, the value of ME remains unclear as well as  a possibility of
linear extrapolation of ME for faint stars.

To clarify the matter, another method of cluster members preselection was used too. 
If we construct the CMD excluding stars with $B-V > 1.0$ and lying far from the most
dense part of the VPD, we can obtain the first approximation of the cluster main
sequence.
Using this criterion, we selected 310 probable cluster members.
They allowed to evaluate
 the ME as $ +0.51 \pm 0.05\;{\rm mas\, yr^{-1}\, mag^{-1}}$ for $\mu_x$,
 and $ -0.35 \pm 0.04\;{\rm mas\, yr^{-1}\, mag^{-1}}$ for $\mu_y$.
These values  confirm the character of the  previous magnitude
restricted determination, but, and this
is extremely important, they show the same character as the ME determined in the
method of statistical parallaxes for all observed magnitude  $8.5\:{\rm mag} < B < 15.5\:{\rm mag}$.

Finally, we tried to estimate the ME by comparing the obtained proper 
motions with absolute ones obtained in the space projects HIPPARCOS and 
Tycho.
However, it is difficult to determine the ME comparing our proper motions 
with these  of HIPPARCOS.
Our catalogue of more than 2000 stellar proper motion determinations in 
the region of NGC\,7243 covering a small field of 6400 square arcminutes, 
contains only five  stars common with HIPPARCOS. 
All of them are in the $8.5-9.8$ magnitude range and a comparison of the
proper motions does 
not permit to make any significant conclusions about the ME.

The proper motions presented in Tycho-2 catalogue are derived using mainly
photographically observed  first epoch positions. 
However, a large number  of stars common with  HIPPARCOS in all observed 
magnitude range, permitted to correct the main part of the principal systematic 
errors, including ME, in  Tycho-2.
In the studied field of NGC\,7243 we found 280 stars common with Tycho-2.
The ME obtained as the result of comparison of proper motions 
with Tycho-2 can be presented as:

    $+0.52 \pm 0.17\;{\rm mas\, yr^{-1}\, mag^{-1}}$ for $\mu_x$,

    $-0.53 \pm 0.22\;{\rm mas\, yr^{-1}\, mag^{-1}}$ for $\mu_y$.

The low relative precision of the obtained ME and, in principal,  the narrow range of 
stellar magnitudes does not permit to apply them directly to 
correct the observed  proper motions but permit to confirm the previously 
determined values of ME.

On the other hand, the realized comparison permits (taking into account 
the errors of our catalog) to estimate the accidental errors 
of Tycho-2 proper motions as:
 $$ \sigma _x = \pm 3.0\;{\rm mas\,yr^{-1}},\;\, \sigma _y =\pm 3.6\;{\rm
 mas\,yr^{-1}}.  $$
Finally, the ME obtained on the base of 310 preliminary photometrically 
selected cluster members was applied for
 every plate pair used for mean proper motion determinations. It was shown, that
 the same ME, within the error bars, may be used for all  plate pairs.
 This circumstance gave us an opportunity to use this ME in our following
 investigations applying the
 corrections caused by the ME to all determined proper motions.

 The procedure for a final selection of cluster members described below
 permitted to create
 the list of probable cluster members. The original (or observed) proper motions
 of these
 probable cluster members were studied for an eventual existence of ME.
 The ME obtained in the second
 approximation shows a linear character with the following regression
 coefficients (for the whole
 magnitude range):

       $+0.49 \pm 0.05\;{\rm  mas\,yr^{-1}\, mag^{-1}}$ for $\mu_x$,

       $-0.17 \pm 0.06\;{\rm mas\,yr^{-1}\, mag^{-1}}$ for $\mu_y$.

 All proper motions derived in our investigation, were corrected using  this
 ME for all observed stellar magnitudes. The resulting ME is shown on Fig.~2.
 The third approximation, carried out following the same scheme, does not show
 any significant residual ME.

\begin{table*}  
 \caption[]{Field stars and the cluster members selected in the previous works}
 \begin{tabular}{ccccc|ccccc}
 \hline
 \multicolumn{4}{c}{Star numbers derived from}&& \multicolumn{4}{c}{Star numbers
     derived from}& \\
  This &  Lenga-&   Sche- & Mumi-& $V_{\rm rad}$ &  This  & Lenga- &  Sche- &
 Mumi- & $V_{\rm rad}$\\
 paper &  uer   &   wick  & nov  & ${\rm km\,sec^{-1}}$    &  paper &  uer   &
 wick  &  nov  & ${\rm km\,sec^{-1}}$ \\
 \hline
  166$\;\,    $& 785\, &        &      & $  +7.4 $     & 891$\;\,    $    & 492*    &   59*  & 632     &          \\
  303$\;\,    $& 735\, &        &      & $ -34.8 $     & 895$^\dagger$    & 362\,   &   57*  &         & $-4.0$   \\
  350$\;\,    $& 754\, &        &      & $ -27.8 $     & 896$\;\,    $    & 361*    &   56*  &  625\,  & $-11.0$ \\
  553$^\dagger$& 649\, &        &      & $ -18.3 $     & 897$\;\,    $    & 363*    &   58\, &  626\,  &   \\
  616$\;\,    $& 500*  &   77\, & 676  &               & 902$^\dagger$    &  358\,  &   49*  &  597\,  & $ -15.0 $\\
  743$\;\,    $& 476*  &   23\, & 475  & $ -52.0 $     & 904$\;\,    $    & 487*    &   52*  &  607\,  & $ -35.0$\\
  744$\;\,    $& 473*  &   13*  & 440  &               & 905$\;\,    $    &  483\,  &   46*  &  586\,  & \\
  757$^\dagger$& 482*  &   42*  & 567  & $  -8.0 $     & 909$\;\,    $    & 480*    &   33*  &  516\,  & $ -11.0 $\\
  759$^\dagger$& 484*  &   44*  & 587  & $ -13.0 $     & 912$^\dagger$    & 475*    &   21*  &  467*   & $ -13.0 $\\
  761$^\dagger$& 485*  &   48\, & 592* & $ -11.0 $     & 922$\;\,    $    & 338*    &    7\, &  393*   & \\
  768$\;\,    $& 493\, &        &      & $  +6.0 $     & 924$^\dagger$    & 471*    &   11*  &  416\,  & $ -7.0 $\\
  771$\;\,    $& 490*  &   57*  & 628  & $  +5.0 $     & 1000$\;\,    $   &  306\,  &        &         & $ +9.2 $ \\
  775$^\dagger$& 497*  &   75\, & 672  & $ -27.0 $     & 1068$\;\,    $   &  364    &   65*  &  642\,  & \\
  776$^\dagger$& 496*  &   71*  & 661  & $  -6.0 $     & 1075$^\dagger$   &   368*  &   83*  &  696\,  & $ -5.0 $\\
  778$\;\,    $& 501*  &   78\, & 682  &               & 1078$^\dagger$   &   370*  &   85*  &  714\,  & $ -12.0 $\\
  780$^\dagger$& 503*  &   80*  & 685  & $  -6.0 $     & 1081$\;\,    $   &   374*  &   91\, &  751\,  & \\
  814$^\dagger$& 527\, &        &      & $ -24.0 $     & 1084$\;\,    $   &   378*  &   96*  &  769\,  & \\
  869$\;\,    $& 382*  &  109\, & 808  &               & 1085$^\dagger$   &   380*  &  102*  &  783\,  & $ -23.0 $\\
  872$^\dagger$& 512*  &   99*  & 785  & $ -18.0 $     & 1086$\;\,    $   &   379\, &  100*  &  782*   & $ -37.0 $  \\
  876$\;\,    $& 510*  &   88*  & 731  & $  +3.0 $     & 1097$^\dagger$   &   384*  &  111*  &  816\,  & $ -20.0 $\\
  879$^\dagger$& 377*  &   94*  & 765  & $ -13.0 $     & 1099$^\dagger$   &   386\, &  114\, &  828\,  & $ -9.0 $\\
  881$^\dagger$& 372\, &   89   & 745  & $  -4.0 $     & 1104$^\dagger$   &   392\, &  125\, &  892\,  & $ -11.0 $\\
  886$^\dagger$& 369\, &   84*  & 699  & $  +4.0 $     & 1737$^\dagger$   &   181\, &        &         & $-14.3 $\\
 \hline
\multicolumn{10}{l}{* - previously determined cluster members,} \\
\multicolumn{10}{l}{$^\dagger$ - probable cluster members based on the radial velocity,} \\
\multicolumn{10}{l}{stars  743, 768, 771, 896, 904, 909, 1085, 
       and 1086 are spectroscopic binaries}
\end{tabular}
\end{table*}

\begin{figure} [t]
  \vspace {7.1cm}
  \caption []{Final magnitude equation. Points present mean proper motions of
           selected cluster members for
              different stellar magnitude groups.
              The analytical ME is obtained as a linear fitting to
              the $\mu _x$ and $\mu _y$ components    }
  \vbox{\includegraphics{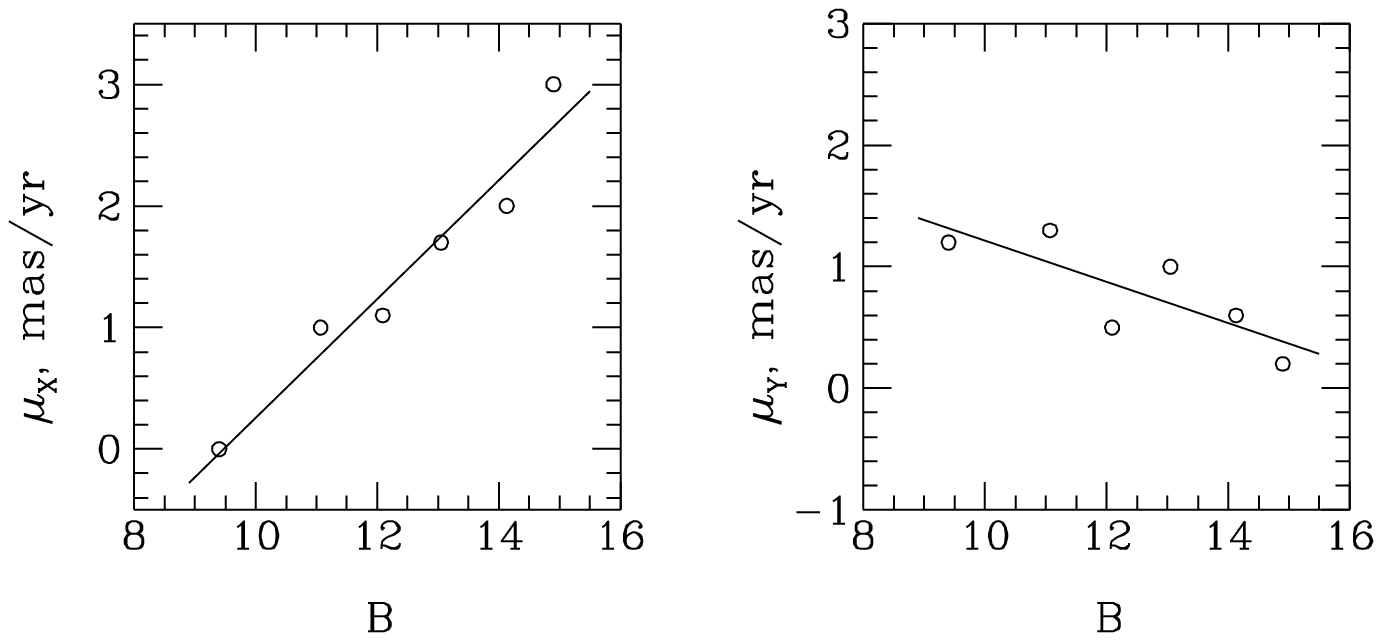}} %-140
 \end{figure}

The cluster member selection was done based on VPD and CMD.
For these purposes, a preliminary
catalogue was constructed including all stars having
originally  determined proper motions and reliably determined $B$ and $V$
magnitudes (in the published version of the catalogue the values without symbol
 ``:'').
In every possible case, photographic photometric data were substituted by
their photoelectric counterparts. The VPD and CM diagram
were constructed for all
1688 stars of this preliminary catalogue (Figs.~3 and~4). Fig.~5
presents an enlarged central part of the VPD.
No condensation  corresponding to the cluster can be seen.

Probable ``astrometric'' cluster members were determined by the formula
proposed by Vasilevskis et al. (1958) %(\cite{vasilevskis})
and
Sanders (1971)      %(\cite{sanders})
based on the idea that proper motion distributions for cluster and
field stars were  bivariate Gaussian distributions, circular for the cluster
and elliptical for the field.
Thus, each point of the VPD takes part in two
Gaussian distributions: that of the cluster $\Phi_c (\mu_x,\mu_y)$ and that
of the field $\Phi_f (\mu_x,\mu_y)$:
%
%\begin{center}
$$ \Phi \left(\mu_{x_i},\mu_{y_i}\right) = \Phi_c
 \left(\mu_{x_i},\mu_{y_i}\right)+
 \Phi_f \left(\mu_{x_i},\mu_{y_i}\right),$$

$$ \Phi_c\left(\mu_{x_i},\mu_{y_i}\right)=
 \frac{N_c}{2\pi\sigma^2}\times
 \exp\left \{-\frac{1}{2}\left[ \left( \frac{\mu_{x_i}
 -\mu_{{x_0}c}}{\sigma}\right)^2 + \left( \frac{ \mu_{y_i}-\mu_{{y_0}c}}
 {\sigma} \right)^2 \right] \right\},\eqno (2) $$

$$ \Phi_f
 \left(\mu_{x_i},\mu_{y_i}\right)=\frac{N_f}{2\pi\Sigma_x\Sigma_y\sqrt{1-r^2}}
 \times
  \exp \left \{ -\frac{1}{2 \left(1-r^2\right) } \left[ \left( \frac{\mu_{x_i}
 - \mu_{{x_0}f}}{\Sigma_x} \right)^2 + \left( \frac{\mu_{y_i}-\mu_{{y_0}f}}
 {\Sigma_y} \right)^2
 -2r\frac{\left(\mu_{x_i}-\mu_{{x_0}f}\right)}{\Sigma_x}
 \frac {\left(\mu_{y_i}-\mu_{{y_0}f}\right)}{\Sigma_y}\right] \right\} $$
%\end{center}
% \bigskip

 where
      $N_c$ - normalized number of cluster stars,

      $N_f$ - normalized number of field stars,

     $\sigma$ - standard deviations for stars of the cluster,

   $\Sigma_x$, $\Sigma_y$ - standard deviations for the field stars,

      $r$ -    correlation coefficient, relating to the orientation
            of the ellipse describing the distribution of field stars,

  $\mu_{{x_0}c}$, $\mu_{{y_0}c}$, $\mu_{{x_0}f}$, $\mu_{{y_0}f}$
       - coordinates of centers of distribution
  on the VPD of the
            cluster and  field stars,

 $\mu_{x_i}$, $\mu_{y_i}$ - proper motion in $x$ and $y$ for the $\mbox{\it
 i}^{\,th}$ star.

 Individual membership probability of a star is determined
  by its position on the VPD and is calculated by the following formula:
$$ P \left(\mu_x,\mu_y\right) = \frac{N_c\Phi_c \left(\mu_x,\mu_y\right)}
 {N_c\Phi_c \left(\mu_x,\mu_y\right) + N_f\Phi_f \left(\mu_x,\mu_y\right)} 
  \eqno  (3)$$
For calculating the membership probabilities it is necessary  to determine the
parameters  of each  distribution. 
For this purpose, Sanders (1971) %(\cite{sanders})
suggested to use  a method of maximum likelihood.
Thus, it is necessary
to solve a system of nonlinear equations, the number of which corresponds
to the number of unknown parameters.

 % FIG 3 *******
\begin{figure} [t]
   \vspace {9.2cm}
   \caption{The VPD for all stars in the region of the Open Cluster NGC$\,$7243}
   \vbox{\includegraphics{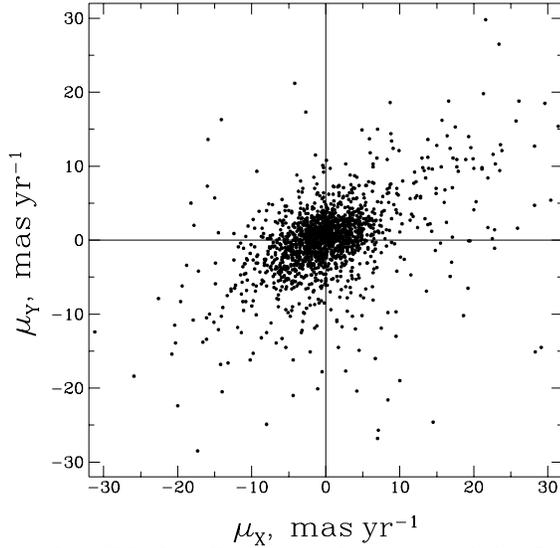}}
\end{figure}

 % FIG 4 *******

\begin{figure} [t]
   \vspace {9cm}
   \caption{CM diagram for all stars in the region of the Open Cluster
    NGC$\,$7243}
   \vbox{\includegraphics{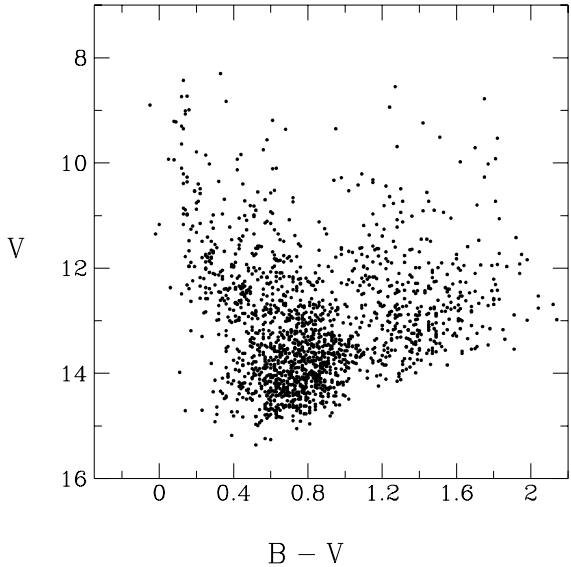}}
\end{figure}

 % FIG 5 *******

\begin{figure} [t]
   \vspace {8.7cm}
   \caption{The central part of the VPD with the orientation of the 
            large axis  of the elliptic proper motions distribution}
   \vbox{\includegraphics{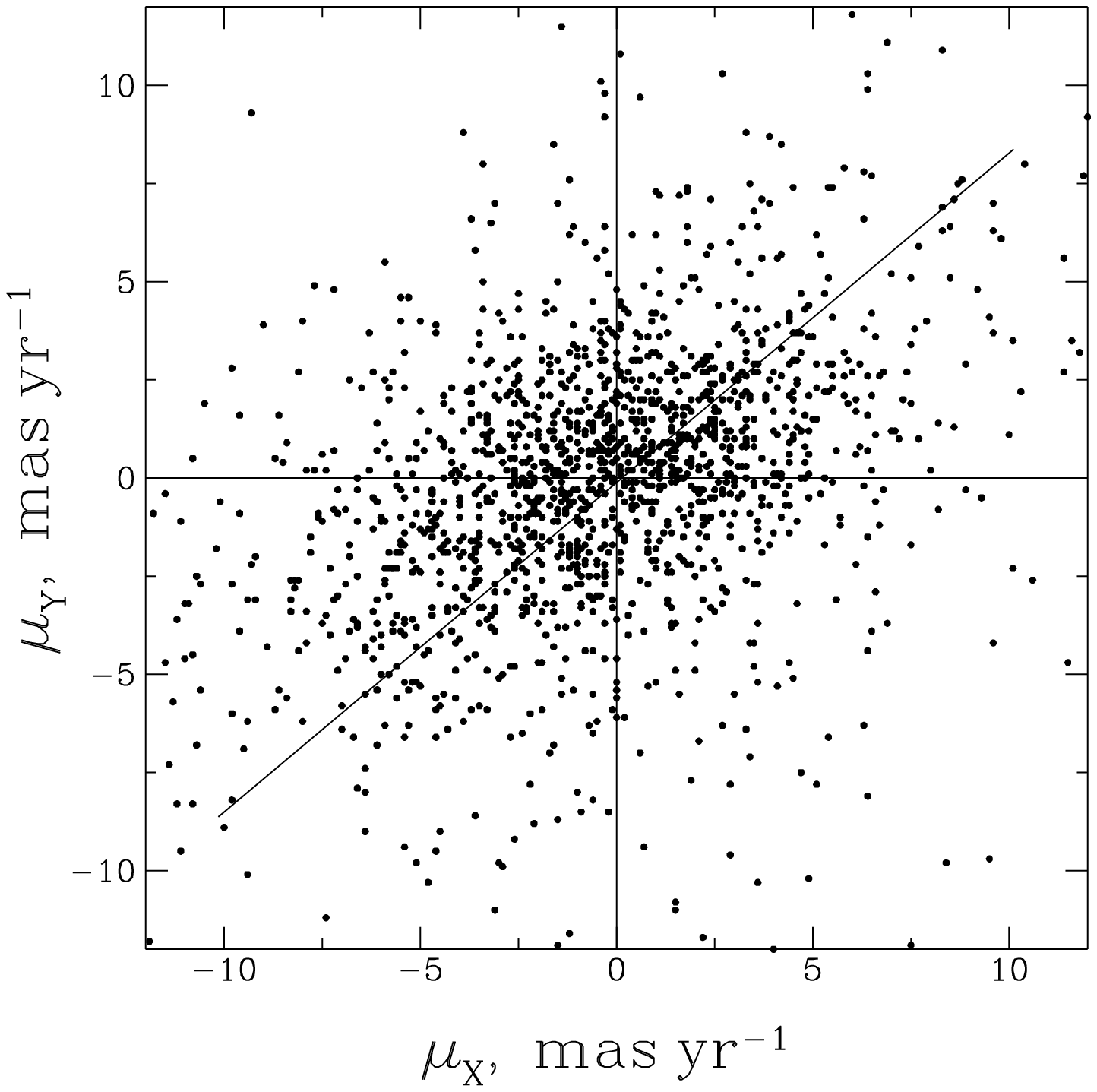}}
\end{figure}

In our study the Sanders' method can not be used directly to fit all 8
unknown parameters of the Gaussian proper motion distributions due to a 
rather small ratio  of cluster and field stars.
The cluster NGC$\,$7243 region of $80\arcmin \times 80\arcmin$
contains more than 2600 stars up to $B=15.5$ mag while according to previous
investigations the cluster diameter
is about $21\arcmin$ (Lyng\aa\, 1987) %\cite{lynga}
and the number of cluster members determined by Lengauer (1933) is only 32.
That is why it was necessary to subject the VPD to preliminary pruning.
The frequency distributions parameters for field and
cluster stars were chosen iteratively.
At the beginning, all originally determined proper motions were transformed
to the new coordinate system ($\mu_{x'},\; \mu_{y'}$)
by rotating the initial coordinate system so
that the large axis of the ellipse coincides with the $x\,$ (Right Ascension)
axis.
The preliminary center of the cluster proper motion distribution was 
determined  in this new coordinate frame.
For a preliminary cluster members separation,  a procedure similar to the
case of ME determination was used.

The visual analysis of the CMD  shows
that the stars with $B - V > 1.0$ are improbable cluster members and might be
excluded
from  consideration. So, the elimination of these stars will clean
the central part of the VPD.
On the other hand, stars with large proper motions
($|\mu_{x'}, \mu_{y'}| > 50\,{\rm mas\,yr^{-1}}$)
might be eliminated from  consideration too.

Taking into account that the cluster diameter is $21\arcmin$ (Lyng\aa\, 1987),
 %\cite{lynga})
stars lying at distances more than $25\arcmin$  from the geometrical center 
of the cluster were eliminated.

We tried to find the center of the cluster proper motion distribution as 
the region of the highest density on cleaned VPD. 
For this purpose, we compared  the proper motions of each star,
$\mu_{x'_i}$ and $\mu_{y'_i}$, with the proper motions of all residual stars 
and counted the number of stars $N_i$ lying within the circle centered 
on $\mu_{x'_i}$ and $\mu_{y'_i}$  with the radius determined by errors of 
proper motion determination ($\pm  1.1\, {\rm mas \,yr^{-1}}$).
After that, the point with the coordinates ($\mu_{x'_i}$, $\mu_{y'_i}$) was
assigned the  weight $W_i = N_i$.
When this comparison was made for all stars of the cleaned VPD we found
the mean weight  $W_0$.
At the second step, all points on the VPD with $W_i < W_0$ were
eliminated as corresponding to improbable cluster members. For all stars with
$W_i > W_0$ the position of cluster center on the VPD was calculated as a 
barycenter:

 \medskip
      $\mu_{x'_0} = \frac{\Sigma \mu_{x'_i} \cdot W_i} {\Sigma W_i}$, \,\,\,\,\,
      $\mu_{y'_0} = \frac{\Sigma \mu_{y'_i} \cdot W_i} {\Sigma W_i}$.
 \medskip

% FIG 6 *******

\begin{figure} [t]
   \vspace {8cm}
   \caption{Membership probability histogram for all stars in the cluster
 region}
   \vbox{\includegraphics{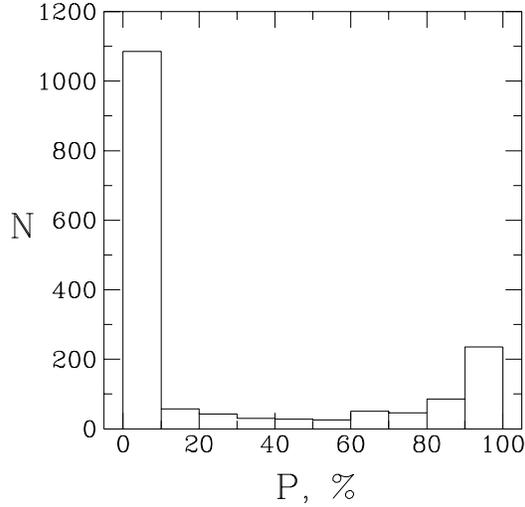}}
\end{figure}

 It is impossible to determine the center of the very noisy proper
 motion distribution of field stars using only VPD, so it was assumed that
 $\mu_{x'}=0$
 and $\mu_{y'}=0$. The ellipse semi-axes or standard deviation of the field
 stars proper motion distribution were determined by VPD.
 The numbers of the cluster members $N_c$ and field stars $N_f$ as well as
 their ratio $N_c/{N_f}=1.03$ were determined from
 this first approximation. Using these values of the distribution parameters,
 individual membership probabilities ($P$) were calculated.

 At the second approximation, only stars with $P> 80\%$ were
 considered as astrometric cluster members. Fig.~6 shows the histogram
of the membership probability.

%***** Fig.7 ***

\begin{figure} [t]
   \vspace{8cm}
   \caption []{Histograms of the proper motions  $\mu_{x'}$ and $\mu_{y'}$
            distributions   and their approximations by fitted Gaussian 
            distributions for cluster and field stars.}
   \vbox{\includegraphics{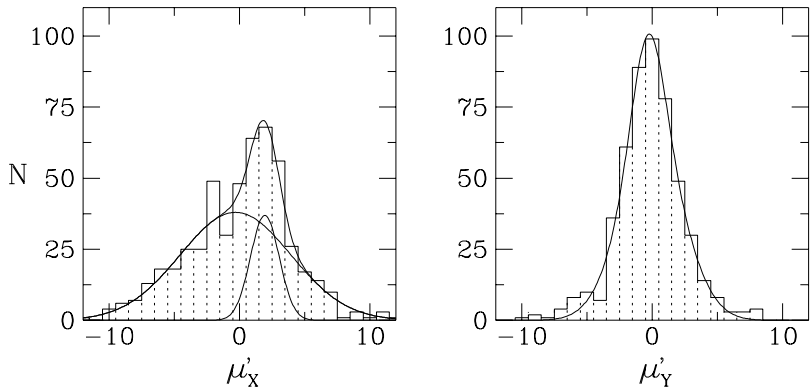}}  %-370
\end{figure}

Then, all stars with membership probabilities $P > 80\%$
were excluded from previously selected field stars, and afterwards
new values of the distribution parameters {\bf of field stars} were obtained.
In this case
Sanders' method was used independently to fit 5 unknowns using only
$\mu_{x'}=0$ and $\mu_{y'}=0$ as the first approximation  of parameters obtained 
from the first iteration.
The resulting values (${\rm mas\,yr^{-1}}$) of the {\bf fitted parameters
of the field and the cluster star distributions are the following}:

\begin{table}[h]
\begin{tabular}{rrrr}

  \underline{The field}           $\mu_{x_0'}$  =     &   $- 1.5$ &
 \underline{The cluster} $\mu_{x'_0}$=  & 2.1\\
              $\mu_{y_0'}$  =     &    0.5 &     $\mu_{y'_0}$=  &  $-1.3$\\
              $\Sigma_{x'}$ =     &    9.0 &     $\sigma_{x',y'}$= &  1.1\\
              $\Sigma_{y'}$ =     &    6.5 &     &      \\
\end{tabular}
\end{table}

The fitted distributions are shown in Fig.~7 where they can be compared with 
histograms of the observed proper motion distributions.

The star position on the cluster CM diagram was accepted as a photometric
criterion of the cluster membership. The shape of the diagram (see Fig.~8)
clearly reflects
the fact that $\sigma_{B-V}$ of $B-V$ determination increases with stellar
magnitude.
All stars with astrometric membership probabilities more than 80\%
and lying in a band of $\pm 3\sigma_{B-V}$ were considered to be
cluster members. The total number of such stars is 209.
Two bright stars, 895 and 896, were added to our list of cluster
 members according to the results of previous determinations of their radial 
velocities and  photoelectric $UBV$ magnitudes. Unfortunately,
 these stars form one non-distinguished image on our plates.
 So the total number of stars which are cluster members is  211.

\begin{figure} [t]
   \vspace{9.0cm} %10.0
   \caption []{CM diagram for probable cluster members with astrometric
            membership probabilities $P>80\%$}
   \vbox{\includegraphics{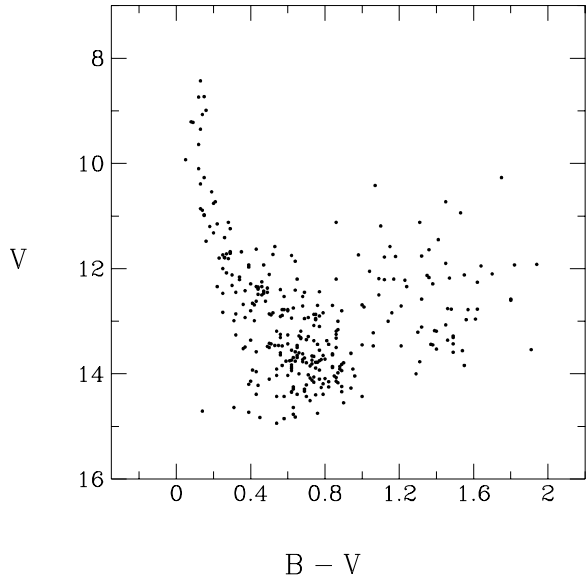}} %-50
\end{figure}

 The comparison of the cluster and the field magnitude functions (Fig.~9) shows
 that the obtained number of cluster members is restricted by the depth
 of the astrometric plates.
From Fig.~10 presenting the XY positions of all the stars in the region of
NGC\,7243, we can see that the field stars are homogeneously distributed over the 
whole area.

\begin{figure} [t]
   \vspace{8.5cm}
   \caption []{The cluster (full line) and the field (dash line) magnitude
 functions}
   \vbox{\includegraphics{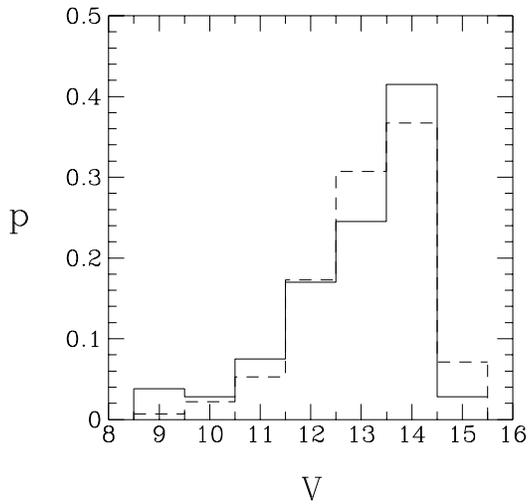}}
\end{figure}

\begin{figure} [t]
   \vspace{7cm}
   \caption []{The position diagram for the NGC$\,$7243 region:
               {\it a)}  all stars in the cluster region, $\;\;\;$
                {\it b)}  selected cluster members,
               {\it c)}  the field stars only}
   \vbox{\includegraphics{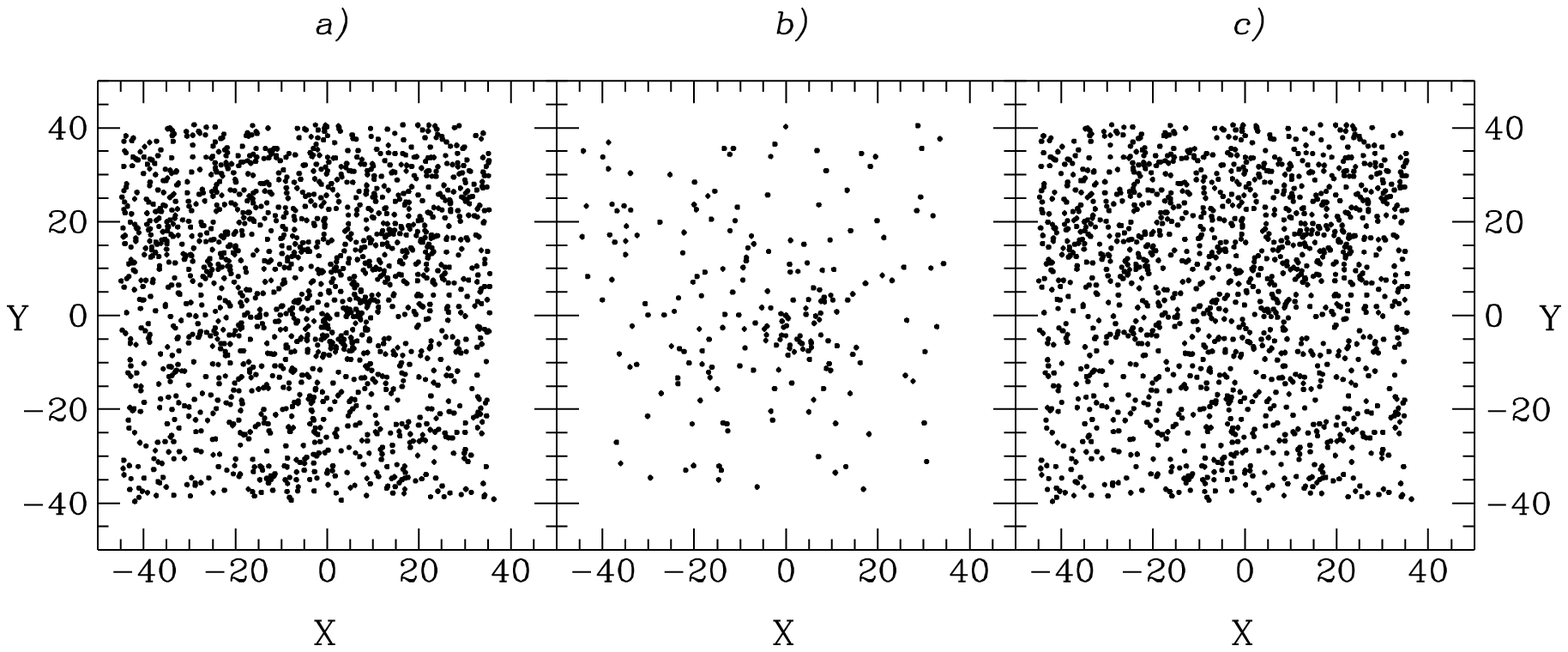}}
\end{figure}

\section{CM diagram and cluster physical parameters}
%______________________________________________________________________________

\subsection*{a) Reddening and distance}

 The selective absorption in the region of the cluster was determined
 using  34 cluster members that satisfied the following conditions:
 {\it 1)} existence of the $UBV$ photoelectric magnitudes, and {\it 2)}
 luminosity classes IV and V.
 The best coincidence was achieved by overlapping the normal colours line
 and the position of stars on the colour-colour diagram at the value of
  $E_{B-V} = 0.24$~mag.
 The standard inclination of a reddening line, $E_{U-B} / E_{B-V} = 0.72$,
 was taken.
 The results are shown in Fig.~11. The value of colour excess for $B-V$
 completely coincides with similar determinations in the works of MD
 and HB. The last authors applied also the data of the intermediate-band $uvbyH\beta$
 photometry and showed that the absorption in the region of the cluster
 core can be considered as uniform.
 HB, based on their spectral MK classification, obtained a slightly different
 value of reddening: $E_{B-V}= 0.19 \pm 0.05$~mag.

\begin{figure} [t]
   \vspace{9cm}
   \caption []{The two colour diagram of the open cluster NGC~7243 approximated
    with a line of normal colours}
   \vbox{\includegraphics{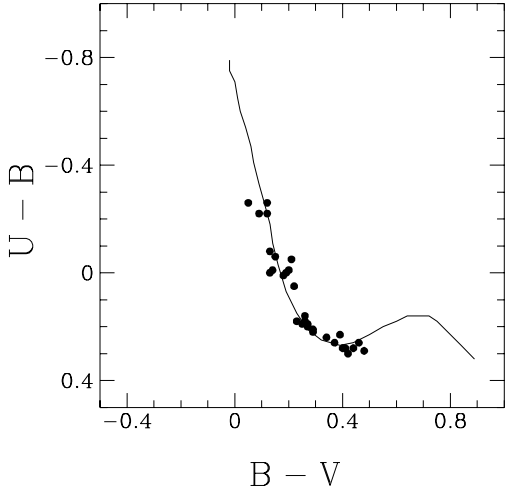}}
\end{figure}

The visible distance modulus  was obtained by a shift of ZAMS
on the CM diagram along the axis $V$ at $E_{B-V} = 0.24$~mag up
to its coincidence with the MS of the cluster. Despite a noticeable scattering
of the points, the distance modulus can be determined due to the fact that the
positions of ten stars, which are highly probable cluster members
(based on their astrometric, photometric and radial velocities 
determinations) form a
visible thickening on the diagram shown by a circle in Fig.~12, where the
final CM diagram  of 211 members of the open cluster
NGC$\,$7243 is plotted. Its middle
point is: $V= 11.74$~mag, $B-V= 0.28$~mag, and the corresponding value of the
visible distance modulus is $V-M_V = 9.94$~mag, that at a general
absorption of $A_V = 0.72$~mag results in the true modulus of
$V_0-M_V = 9.22$~mag.
The points of the lines of normal colours and ZAMS were taken from the work of
Mermilliod (1981)    %(\cite{mermilliod})
where they were determined using open clusters younger than Hyades.
The determination of the average values of visible modulus for astrometric
members by MK classification results in the value of $V- M_V = 9.94
  \pm 0.37\;{\rm mag}$  that corresponds to a distance of 
$r =  {698^{\,+130}}\!\!\!\!\!\!\!\!\!\!\!_{-109}\,$pc.

\begin{figure} [t]
   \vspace{9cm}
   \caption []{CM diagram for the open cluster NGC~7243 members. The circle
 marks the positions of ten stars, which are reliable cluster members. 
    ZAMS is shown by a solid line.}
   \vbox{\includegraphics{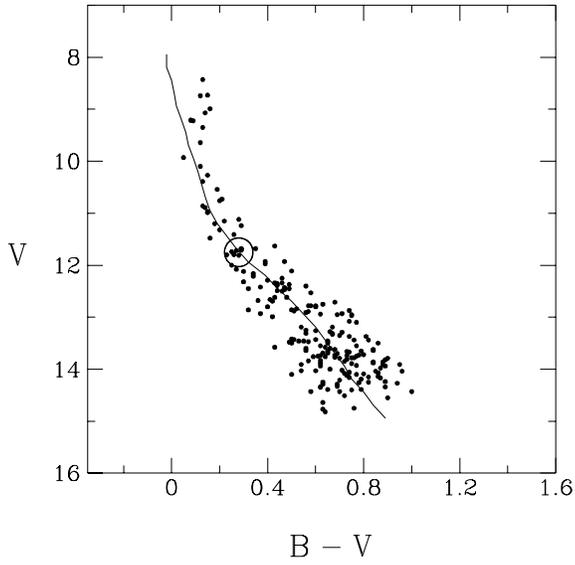}}
\end{figure}
%______________________________________________________________________________

\subsection*{b) Cluster age}

 The age of the cluster was estimated based on the $(B-V)_0$ value of the
 turn-off
 point. This value for NGC\,7243 is
 $(B-V)_0 = - 0.11\;{\rm mag}$. For such data the cluster age was estimated
 as $(2.5\pm0.5)\cdot 10^8\,$yr by the use of the isochrone grids computed by
 the Padova theoretical group (Bertelli et al. 1990). 
 %\cite{bertelli}).
 The main sequence and three isochrones corresponding  to the ages of
 $1\cdot 10^8$, $2 \cdot 10^8$, and $3 \cdot 10^8$ yrs are shown in Fig.~13.

\begin{figure} [t]
   \vspace{10cm}
   \caption []{The CM diagram with ZAMS and three isochrones corresponding to
  different cluster ages}
   \vbox{\includegraphics{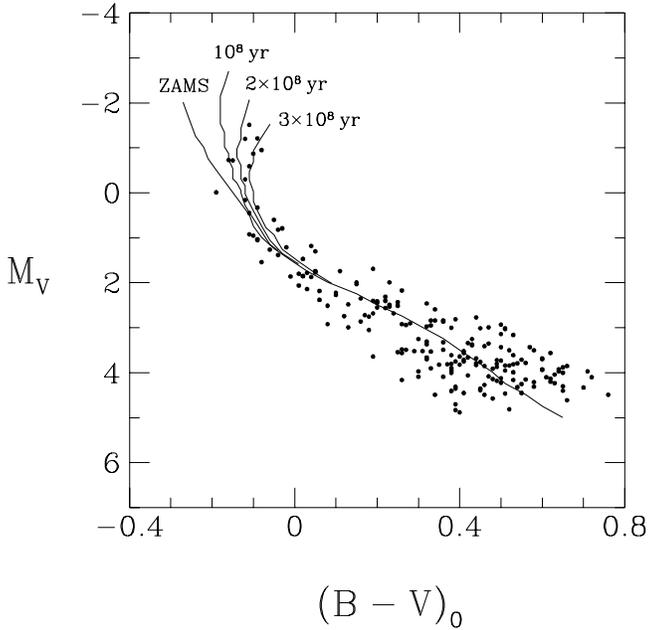}}
\end{figure}

 \subsection*{c) Luminosity and mass functions. Cluster mass estimation}

   The cluster luminosity function, $\phi ( M_V )$, is determined by the
 following equation:  $dN = \phi(M_V)dM$, where $dN$ - number of stars
 with absolute magnitudes in the limits $M_V$, $M_V + dM_V$.
 The normalized luminosity function for NGC$\,$7243 based on our data
 is given in Fig.~14.

 Star masses were derived from the  ``mass-luminosity'' relations
 published by Straizys and Kuriliene (1981).       %(\cite{straizys}).
 Then, the empirical mass function for NGC\,7243 was constructed.
 Salpeter (1955)       %(\cite{salpeter})
 was the first who showed that for the stars
 in the vicinity of the Sun the power law of the initial mass
 function (IMF) has the form of a power law
 $dN/dM=(M/M_\odot)^\alpha$, whith $\alpha=-2.35$.
 Numerous recent
 investigations of star clusters and associations of the Milky Way and
 the LMC exhibit a large range of logarithmic slopes in any mass range.
 Scalo (1998),     %(\cite{scalo})
 assembling and analyzing these data, proposed a
 three-segment power law IMF:

 $$ dN/dM = \left\{ \begin{array}{rcl}
 {-1.2 \pm 0.3} & \mbox{\rm for}  & 0.1 < M/M_\odot <  1  \\
 {-2.7 \pm 0.5} & \mbox{\rm for}  & 1\;\;   < M/M_\odot < 10  \\
 {-2.3 \pm 0.5} & \mbox{\rm for}  & 10  < M/M_\odot < 100 \\
 \end{array} \right.
 $$

 In our investigation the logarithmic mass function (Fig.~15) was obtained
 as rms fitting for stars in the mass range $1 < M/M_\odot <6.5 $ in the form

 $$    \lg N=-2.37 \lg (M / M_\odot)+2.18.  $$

 Thus the  inclination of the mass function of NGC$\,$7243 $(-2.37\pm0.27)$
 is in a good agreement with $\alpha = -2.3$ recently obtained by Meusinger
 et al. (1996)      %(\cite{meusinger})
 for stars with $1.1 < M/M_\odot < 3.5$ of the Pleiades cluster.

 \begin{figure} [t]
   \vspace{8.5cm}
   \caption []{The normalized luminosity function for the open cluster NGC~7243}
   \vbox{\includegraphics{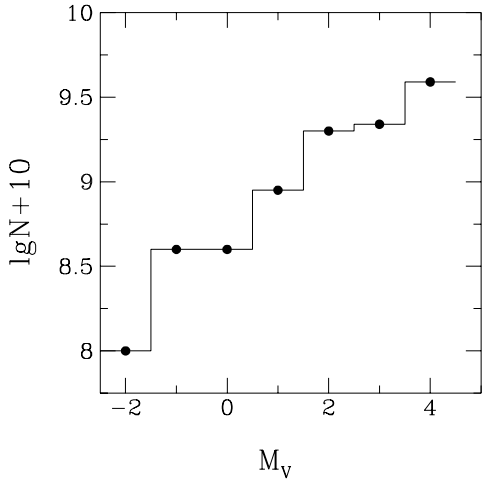}}
 \end{figure}

 \begin{figure} [t]
   \vspace{8cm}
   \caption []{Mass function of the open cluster NGC~7243}
   \vbox{\includegraphics{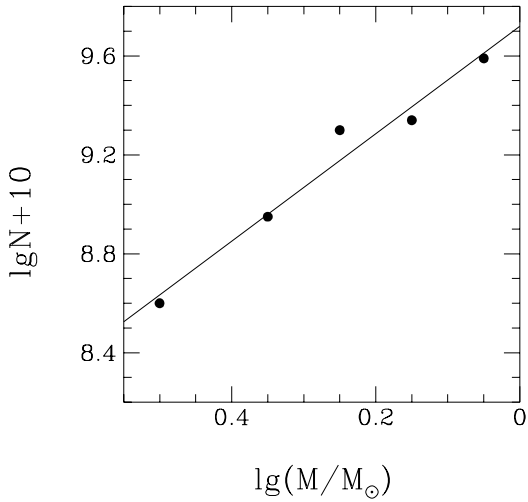}}
 \end{figure}

 The estimation of the cluster mass was made on the following suggestions.
 The direct determinations of masses of each of its members based on the
 mass-luminosity correlation  give  a total value of $348M_{\odot}$.
 Assuming that a half of them are double stars with a ratio of masses of
 components close to unity, we should increase this value in $1.5$ times.
 The resulting estimation of $522M_\odot$ is still not final since it is
 possible that part of the faint cluster members has remained unrevealed
 because of the depth limit  of the astrometric plates.

\subsection*{d) Variable and peculiar stars in the region of the cluster}

     According to the General Catalogue of Variable Stars (1985)
 %(\cite{kholopov1})
 and the
 New Catalogue of Suspected Variable Stars (1982)     %(\cite{kholopov2})
 there are 8 known and
 3 suspected variable stars in the studied region. Four of them are
 identified in our catalogue (see Table 8, column of the notes) but
 none of them  are cluster members. Among the suspected variables
 two are not members  and the third - number 909 in our catalogue (number 480
 in Lengauer one) - is one of the brightest stars of the NGC$\,$7243, spectral
 binary
 with attributes of weak brightness fluctuations according to HB
 observations.
 Our photometry does not allow to prove or disprove this fact, so
 the issue remains open. On the other hand, three other stars from
 our catalogue (863, 1634 and 2278) have shown significant
 divergences in $B$ and $V$ magnitudes obtained with the different plates
 of Schmidt telescope.  The star  863 has a high membership probability
 by proper motions - $P=97\%$.  According to this criterion among the cluster
 members there are also two stars with unusual spectra: number 904 - A${\rm m}$
 with  $P= 85\%$ and 1078 - A0$\,$V${\rm p}$ with $P=96\%$.

 %______________________________________________________________________________

 \section{Conclusions}

We have carried out astrometric and photometric study of the
 young  open cluster  NGC\,7243  poorly investigated before.  
The main  cluster parameters obtained in this investigation are given in Table~7.
The complete catalogue of 2623 stars in the region of the NGC$\,$7243
containing positions, proper motions and their individual errors, $B$ and $B-V$ 
magnitudes and individual membership probabilities (Table 8) is available in 
electronic form at the CDS via anonymous ftp
130.79.128.5.

 \bigskip

 \begin{table*}
     \caption{NGC$\,$7243 summary data}
 \begin{center}
 \begin{tabular}{ll}
 \bigskip

        Number of members              &    $211$\\
        Limiting magnitude             &    $V \sim 15.5\;{\rm mag}$\\
        Magnitude of the brightest star &   $V = 8.43\;{\rm mag} $\\
        Earliest spectral type         & B5$\,$III \\
        MS turn-off point              & $(B-V)_0 = - 0.11\;{\rm mag}$\\
        Age                            & $t=(2.5\pm 0.5)\cdot 10^8\,$yr\\
        Total cluster mass    & $348M_{\odot} \le M_{TOT}\le 522M_\odot$ \\
        Average reddening     & $E_{B-V}=0.24\;{\rm mag}$\\
        Distance              &  $r =  {698^{\,+130}}\!\!\!\!\!\!\!\!\!\!\!_{-109}\,$pc \\
        Mass function slope            & $-2.37 \pm 0.27$  \\
 \end{tabular}
 \end{center}
 \end{table*}

\begin{table*}   
\begin{center}
        \caption[]{Complete catalogue of 2623 stars in the region of the NGC\,7243 (example of the Table~8 
                   only available in electronic form at the CDS)}
\begin{tabular}{rccccccccccccccc}
\hline
 This &RS&Len-& $\alpha$   &   $\delta$   & Pair & $\mu_x$& $\epsilon_x$ & $\mu_y$&$\epsilon_y$& $V$&$B-V$& $P$&  $X$&$Y$& Notes \\
work  &  &gauer & \multicolumn{2}{c}{2000.0} &      & \multicolumn{4}{c}{${\rm mas\,
 yr^{-1}}$}          &    &    &  \%   &   mm    & mm &\\
\hline
   1 & & & 22 10 42.86 &  49 17 39.2  &   9 &    1.7 &  1.0 &    0.9 &   1.2 &   11.45: &    1.20: &   97 & -43.67 &  -36.14 & \\  
   2 & & &22 10 46.88 &  49 16 44.2  &   9 &    1.0 &  1.7 &   -2.3 &   1.6 &   13.40  &    0.77  &   62 & -43.02 &  -37.07 &\\  
   3 & & &22 10 48.33 &  49 19 02.6  &   9 &    8.4 &  1.0 &   -9.8 &   0.8 &   11.78  &    0.74  &    0 & -42.75 &  -34.75&\\ 
   4 & & &22 10 49.61 &  49 14 59.1  &   9 &   -2.1 &  1.2 &   -0.7 &   0.8 &   11.98: &    0.40: &    1 & -42.60 &  -38.85& \\ 
...&...&...&...       &...           &...  &...     &...   &...     &...    &...       &...       &...   &...     &...     & \\
 289 & &717& 22 15 43.11&49 26 57.4  &  10 &     3.7 &  1.0&    1.9 &   0.7 &    8.71: &    0.14:&    89 &   5.64 &  -27.13& HD\,211418 \\
... &... &... &... &... &... &... &... &... &... &... &... &... &... &...  \\
2623&&&  22 10 34.29  &  50 29 02.3  &  10  &    0.9  &  0.7  &   -0.9 &   0.8  &  12.40: &    0.57: &   91 &  -43.93 &   35.79 \\
 \hline
 &&&&&&&&&&\\
 \multicolumn{14}{l}{\small Column (1) star number in our catalogue,} \\
 \multicolumn{14}{l}{\small Column (2) the reference stars are marked with *,} \\
 \multicolumn{14}{l}{\small Column (3) star number according to Lengauer,} \\
 \multicolumn{14}{l}{\small Columns (4) and (5)  equatorial coordinates at
      2000.0,}\\
 \multicolumn{14}{l}{\small Column (6)  number of used astroplate pairs,}\\
 \multicolumn{14}{l}{\small Columns (7) and (8)  proper motion $\mu_x$ and its
      error  $\epsilon_x$,}\\
 \multicolumn{14}{l}{\small Columns (9) and (10) proper motion $\mu_y$ and its
      error  $\epsilon_y$,}\\
 \multicolumn{14}{l}{\small Column (11)  $V$ magnitude,}\\
 \multicolumn{14}{l}{\small Column (12) $B-V$ magnitude,}\\
 \multicolumn{14}{l}{\small Column (13)  value of membership probability $P$,}\\
 \multicolumn{14}{l}{\small Column (14) and (15) rectangular coordinates (relative to the star 898)
            in the scale of the Normal Astrograph,}\\
 \multicolumn{14}{l}{\small Column (16) notes}\\
 \end{tabular}
 \end{center}
 \end{table*}

 \begin{acknowledgements}
  E.G.J. thanks CNPq and FAPERJ for  financial support under contracts
  300016/93-0 and E-26/152.221/2000 N.A.D. thanks FAPERJ for  financial support
  under  contracts E-26/151.172/98 and E-26/171.647/99.
 \end{acknowledgements}

 {}

 \end{document}